\documentclass[10pt]{article}
\usepackage[letterpaper]{geometry}
\usepackage{hicss51}
\usepackage{times}
\usepackage[none]{hyphenat}
\usepackage{url}
\usepackage{latexsym}
\usepackage{indentfirst}
\usepackage{graphicx}
\graphicspath{{images/}}
\usepackage{xcolor}
\usepackage{amsmath}
\usepackage{float}
\usepackage{subcaption}
\title{Fast Solutions in Power System Simulation through Coupling with Data-Driven Power Flow Models for Voltage Estimation}

\author{
 Siobhan Powell \\
 Stanford University \\
 {\underline{ siobhan.powell@stanford.edu}} \\\And
 Alyona Ivanova \\
 SLAC National Accelerator Lab \\
 {\underline{ aivanova@slac.stanford.edu} }\\\And 
 David Chassin \\
 SLAC National Accelerator Lab \\
 {\underline{dchassin@slac.stanford.edu}} \\
}

\date{}

\begin{document}

\maketitle

\begin{abstract}
Power systems solvers are vital tools in planning, operating, and optimizing electrical distribution networks. The current generation of solvers employ computationally expensive iterative methods to compute sequential solutions. To accelerate these simulations, this paper proposes a novel method that replaces the physics-based solvers with data-driven models for many steps of the simulation. In this method, computationally inexpensive data-driven models learn from training data generated by the power flow solver and are used to predict system solutions. Clustering is used to build a separate model for each operating mode of the system. Heuristic methods are developed to choose between the model and solver at each step, managing the trade-off between error and speed. For the IEEE 123 bus test system this methodology is shown to reduce simulation time for a typical quasi-steady state time-series simulation by avoiding the solver for 86.7\% of test samples, achieving a median prediction error of 0.049\%.

\end{abstract}

\section{Introduction}

With electricity demand on the rise and increased diversification of generation sources in the field, sophisticated power system modeling tools are required for electrical utility companies to be able to provide a clean, efficient, and reliable network to all customers. One of the challenges with modeling electrical distribution networks is the growing variety of system modes observed due to the emergence of new distributed resource technologies, making it computationally expensive for traditional power system solvers to produce results during long simulations. In many network design and analysis problems thousands of power flow solutions must be obtained in a deterministic time-sequence referred to as ``quasi-steady state'' to generate data for multiple days of operation or to conduct a sensitivity analysis on different model parameters. Improving the computational performance of quasi-steady state simulations can lead to faster progress in studying and deploying new system control strategies, power system optimization techniques and electricity markets analysis, particularly in the presence of highly variable distributed energy resources. 

Among the open source solvers used to solve quasi-steady state power flow problems are GridLAB-D and Pandapower \cite{Thurner2018PandapowerSystems, Chassin2008GridLAB-D:Environment, Chassin2016NestedGrid}. GridLAB-D in particular has a distinct use-case for power flow solvers, which can be used to compute large scale models that mimic real-world scenarios and aid in decision making processes that address operation, planning and resilience of distribution systems. 

GridLAB-D uses a three-phase unbalanced power flow solver which makes a distinction between radial and non-radial distribution networks. GridLAB-D's default method is forward-back sweep \cite{Kersting2012DistributionAnalysis} for strictly radial systems, while a Newton-Raphson (NR) method is used for non-radial three-phase unbalanced networks. The advantage of the NR method is that it can be applied independent of system-structure and it can provide faster, more reliable results. These advantages come at the cost of increased code complexity and memory usage compared to alternative methods, such as the Gauss-Seidel method \cite{Tinney1967PowerMethod, Schneider2009ModernReport}. The NR methodology calculates voltage updates and current injections at each node by computing the inverse Jacobian of the entire system. Even though the NR methodology is very robust, for large system continuous time simulations reconstructing the Jacobian whenever the system changes significantly can be computationally expensive, and each additional iteration of the solver heavily prolongs the duration of the overall simulation.

GridLAB-D runs time-series simulations that require very large numbers of sequential power flow solutions for conditions that are often similar to those encountered earlier in the simulation, and very often almost identical to the last solution found. To exploit this GridLAB-D uses the most recent simulation output to seed the next solution and reduce the number of iterations required. However, Jacobian updates for states that are close to previously solved states may be computationally redundant, which motivates the development of techniques complementary to the built-in methods that can address these situations. 

Data-driven approximations of the system should be considered as a potentially faster alternative. The idea of using data-driven models to supplement or replace a slow, physics-based simulation has been studied before in other fields including robotics and fluids \cite{Reinhart2017HybridControl, Ladicky2015Data-drivenForests}. In power system data-driven models have been studied for the estimation of voltage, network state, and network parameters \cite{Weng2013HistoricalSystems, Yu2017RobustGrids, Yu2018PaToPa:Grids}.

In this paper we propose the integration of a data-driven power flow model into the NR power flow solution method used by GridLAB-D.
We propose to train a data-driven model on the outputs from the power flow solver, a training data set, and then use the model's estimates to avoid running the full NR solver when it is not necessary, i.e., when the new system state input is sufficiently similar to the training set and confidence in the model's prediction is high.

To fully exploit the potential speed of this approach we implement a model based on Linear Regression, which is faster to train and implement than the Support Vector Regression model detailed in \cite{Yu2017RobustGrids}. Augmenting the Linear Regression approach by learning a new model for each mode of the system (a cluster of similar states in time) we find the estimation is very good for a large majority of samples. We investigate how different fail-safes can be put in place to make the overall model robust and help the simulation choose reliably between the data-driven model solution and the full solver. This is shown to give a large reduction in the number of full NR solves required without compromising the accuracy of the simulation outputs.

This paper is organized as follows: in Section \ref{sec:methods} the model is detailed; in Section \ref{sec:exp} the data set and experiments are outlined,  including the proposed system for coupling the power flow solver with the model; in Section \ref{sec:results} the results of these experiments are presented; they are discussed in Section \ref{sec:disc}; in Section \ref{sec:future} we detail ideas for future testing and improvement of the methodology; and in Section \ref{sec:conc} we present our conclusions. 

\section{Methods} \label{sec:methods}

A load-driven power flow simulation environment calculates the voltage at each bus or measurement point in the system. Let $\{\mathbf{p}, \mathbf{q}, \mathbf{v}, \mathbf{a}\}$ denote the vectors of real power injection, reactive power injection, voltage magnitude, and voltage phase angle at each point in the system where they are measured. There may be a different number of measurement points for the loads and voltages, so let $n_p$ and $n_v$ denote the lengths of the $\{\mathbf{p}, \mathbf{q}\}$ and $\{\mathbf{v}, \mathbf{a}\}$ vectors respectively. 

The mapping being resolved by the solver then is
\begin{align}
    \mathbf{v} &= f_1(\mathbf{p}, \mathbf{q}; Y)\text{,} \\
    \mathbf{a} &= f_2(\mathbf{p}, \mathbf{q}; Y)\text{,}
\end{align}
where $Y$ is the admittance matrix for the network which includes connectivity and line parameter information. 

A data-driven version of this mapping takes the form
\begin{align}
    \mathbf{v} &= \tilde{f_1}(\mathbf{p}, \mathbf{q}) \text{,}\\
    \mathbf{a} &= \tilde{f_2}(\mathbf{p}, \mathbf{q}) \text{,}
\end{align}
where the models $\tilde{f_1}$ and $\tilde{f_2}$ are learned only from historical measurements of $\{\mathbf{p}, \mathbf{q}, \mathbf{v}, \mathbf{a}\}$. We call each measurement a time stamp or sample and denote individual sample measurements with $\mathbf{p}_t$, where $t$ represents time. 

\subsection{Base Model}

In \cite{Yu2017RobustGrids} the authors constructed their data-driven model to match the form of the forward power flow equations, so that learning the model parameters represented learning the values in $Y$. In our experiments we find that constructing the model as a simple Linear Regression gives as good results for the majority of inputs, is simpler to implement in a complex software set-up like GridLAB-D, and is faster to train and apply. For our application the speed is most important and less accurate predictions can be covered by using the real solver. Therefore we formulate $\tilde{f_1}$ and $\tilde{f_2}$ as Linear Regressions
\begin{align}
    \mathbf{v} &= A_1 \begin{bmatrix}\mathbf{p} \\ \mathbf{q}\end{bmatrix} \text{,} \\
    \mathbf{a} &= A_2 \begin{bmatrix}\mathbf{p} \\ \mathbf{q}\end{bmatrix} \text{,}
\end{align}
where $A_1$ and $A_2$ are matrices with dimension $[ n_v, n_p ]$. Given a time-series set of training data, this mapping is learned using the Linear Regression model in the Python sklearn library which implements ordinary least squares \cite{article}. 
\subsection{Clustering}
Testing on a multi-day data set revealed that the system operates in different modes, as the training and testing prediction errors clearly showed time-dependence throughout each day. Clustering in time is implemented to reflect this data pattern in the model. 

The training set is used to sort the time stamps into $n_c$ clusters by their input vectors $[\mathbf{p}, \mathbf{q}]^T_t$. Then one model is trained for each cluster, with each of the $n_c$ clusters becoming a new, smaller training set. Each time stamp in the testing set is sorted into its closest cluster, and the estimate for that time stamp is made using the cluster's model.

Three clustering approaches were considered: first,  we considered K-Means clustering for its versatility and scalability \cite{article, Arthur2007K-means++:Seeding}; second, we considered Gaussian Mixture Models as an extension of K-Means \cite{article, McNicholas2017MixtureClassification}; and third, we adjusted the clusters manually to match the day of the week (7 clusters total). 

\section{Experiment} \label{sec:exp}

An IEEE test network was used to demonstrate the approach, with data generated using GridLAB-D. This section details the experimental set-up, including how the model interacts with the data and power flow solver. The error metric used throughout the results is introduced, and three heuristics for managing the model error are given.


\subsection{Data}

The data set was generated using GridLAB-D, an open-source power flow simulation software \cite{Chassin2016NestedGrid}. The tool is capable of simulating a distribution system on a detailed level in high temporal resolution. Two primary objects, nodes and links, form the majority of the distribution network model. The power flow module within GridLAB-D solves the steady-state node voltages and line currents at each point of the model. The solution of the power flow solver is determined by a three-phase unbalanced flow solver using the NR method \cite{Fuerte-Esquivel2002DiscussionReply}. 

To implement the methodology described in Section \ref{sec:methods}, an IEEE 123 bus standard feeder test network was formulated in the GridLAB-D modeling environment. The feeder model, as described in detail in \cite{DistributionSystemAnalysisSubcommittee1992IEEEFeeder}, operates at 4.16 kV nominal voltage and connects to 344 typical residential houses which were customized to replace the original constant current, power and impedance spot loads. In addition, the circuit accommodates overhead and underground lines, four voltage regulators, shunt capacitor banks and a number of switching elements for controlling two-phase and three-phase lateral feeders.

Coupling the GridLAB-D power flow solver with the IEEE 123 model, we were able to generate a data set to serve as an input to the data-driven power flow solvers described in Section \ref{sec:methods}. The data set was generated using one minute time-resolution, and contains real and reactive power components at the load level and voltage magnitudes and phases at each node for each time step.

\begin{figure}[thb]
	\centering
    \includegraphics[width=\columnwidth]{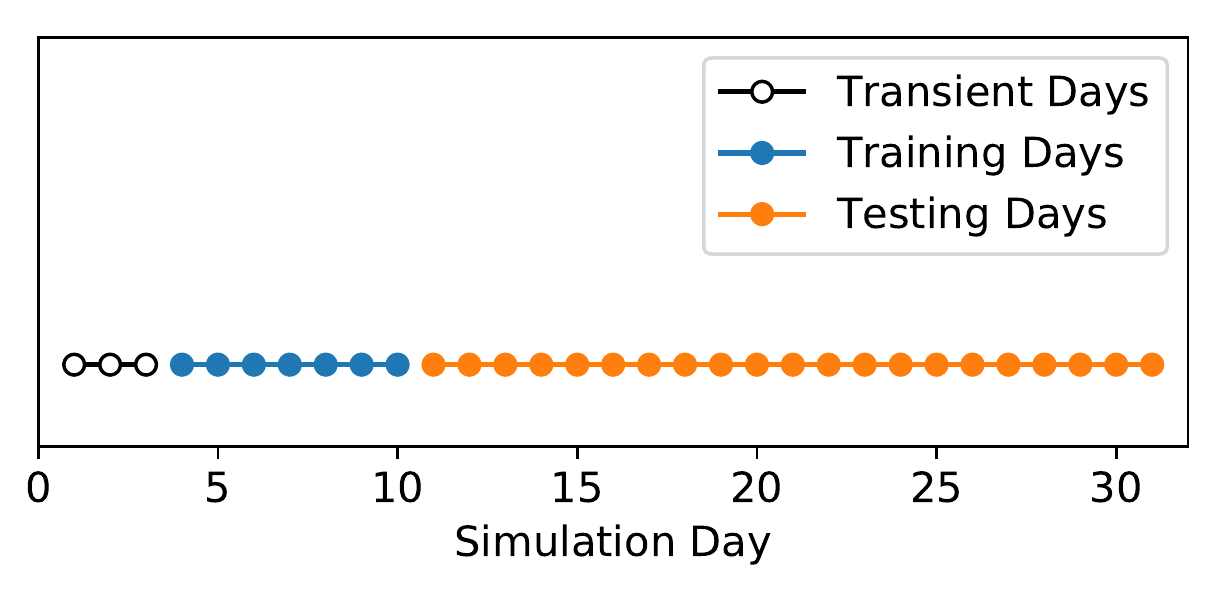}
	\caption{Split of training and testing sets in the month-long data set, including the three dropped days of simulation with initialization transients. }
	\label{fig:datasplit}
\end{figure}

Of the month-long data set created, the first three days were discarded due to the initialization transient behavior of the quasi-steady state simulation, and one week was used for training. A typical week of the data should include all frequently occurring network states and is representative of the future inputs the model will encounter. Since the purpose of this project is also to reduce simulation time, using more than the minimum representative set for training would reduce the utility of this approach. Figure \ref{fig:datasplit} shows how this split was made. The testing set was used as a ``ground truth'' to evaluate the results of the model-integrated approach.

\subsection{Set-Up}

\begin{figure}[thb]
    \centering
    \includegraphics[width=\columnwidth]{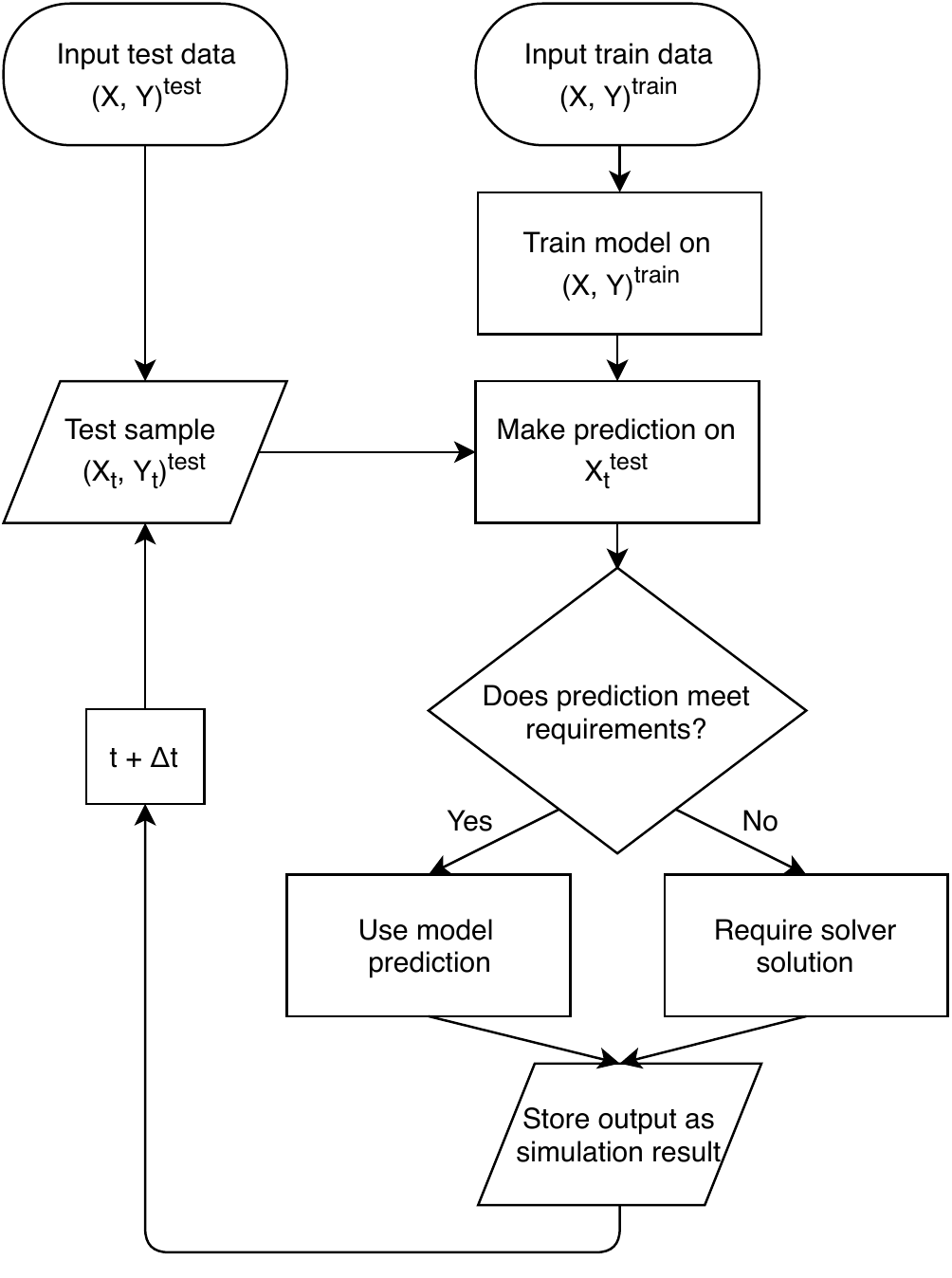}
    \caption{Flow diagram showing the process of building the model, making estimations on testing set samples, and interacting with the full solver to produce the overall simulation result.}
    \label{fig:flow1}
\end{figure}

The main methodology of the test and the interactions of the solver are shown in Figure \ref{fig:flow1}.

In the experiments this process is followed for one day, one week, or all three weeks of the test set. The end output is a mix of estimates from the model and results from the full solver. Comparing this output to the true solution gives an estimate of the error for this approach. 

\subsection{Error Metric}
In this approach we build a model for $\mathbf{v}$ and $\mathbf{a}$ separately, but to represent the error well we use a metric of the total vector error. Where $v$ and $a$ are the values at an individual bus, $\vec{v} = v \exp{(ia)} = [v \cos{a},~ v \sin{a}]$. We predict this vector for each of the $n_v$ voltage measurement points, $\tilde{\vec{v}}$, and construct the normalized vector error: 
\begin{equation}
    \epsilon_{\vec{v}} = \frac{||\tilde{\vec{v}} - \vec{v}||_2}{||\vec{v}||_2} \text{.}
\end{equation}
No normalization is done where $||\vec{v}||_2$ is zero. Doing this calculation for each measurement point gives a $n_v$ vector of errors, $\epsilon_{\vec{\mathbf{v}}}$, and we calculate a measure of the overall prediction error as
\begin{equation}
    \epsilon_{\text{inf} \vec{\mathbf{v}}} = || \epsilon_{\vec{\mathbf{v}}} ||_\infty \text{.}
\end{equation}
We use this error metric throughout the results of this paper, and every time errors are given they are values of $\epsilon_{\text{inf} \vec{\mathbf{v}}}$.

\subsection{Error Parameters}

We consider three tools for making this method more robust and protecting against high-error predictions: 
\begin{enumerate}
    \item \textbf{Error Check.} When the solver is used, apply the model as well and store the error of the estimate. Require that this estimate be below a threshold to invoke the model again. 
    \item \textbf{Distance Check.} Some clusters do not perform as well as others when evaluated on the training set. For under-performing clusters where difficult prediction points have been grouped together, the higher errors typically occur when the sample is near the edge of the cluster. Calculate in what percentile of distance away from the cluster center the new data point falls, and require that this distance be below a certain percentile to use the model. 
    \item \textbf{Step Change Check.} When there is a significant jump between the model's current estimate and the previous simulation solution, that can be an indicator that the model prediction is about to deviate from the true solution. Require that this change in time be below a certain threshold to use the model.
\end{enumerate}

Each of these methods involves parameter tuning and poses a trade-off: making more conservative choices will avoid occasional high error estimates but also slow down the simulation by using the solver more than necessary. 



\section{Results} \label{sec:results}

Following the methods and experimental set-up described above, we carry out tests to select the best model and error parameters for this data set. Using those modeling choices we then outline results showing the performance relative to the original solver solution.  

\subsection{Clustering and Model Choice} 

\begin{figure}[thb]
    \centering
    \includegraphics [width=\columnwidth] {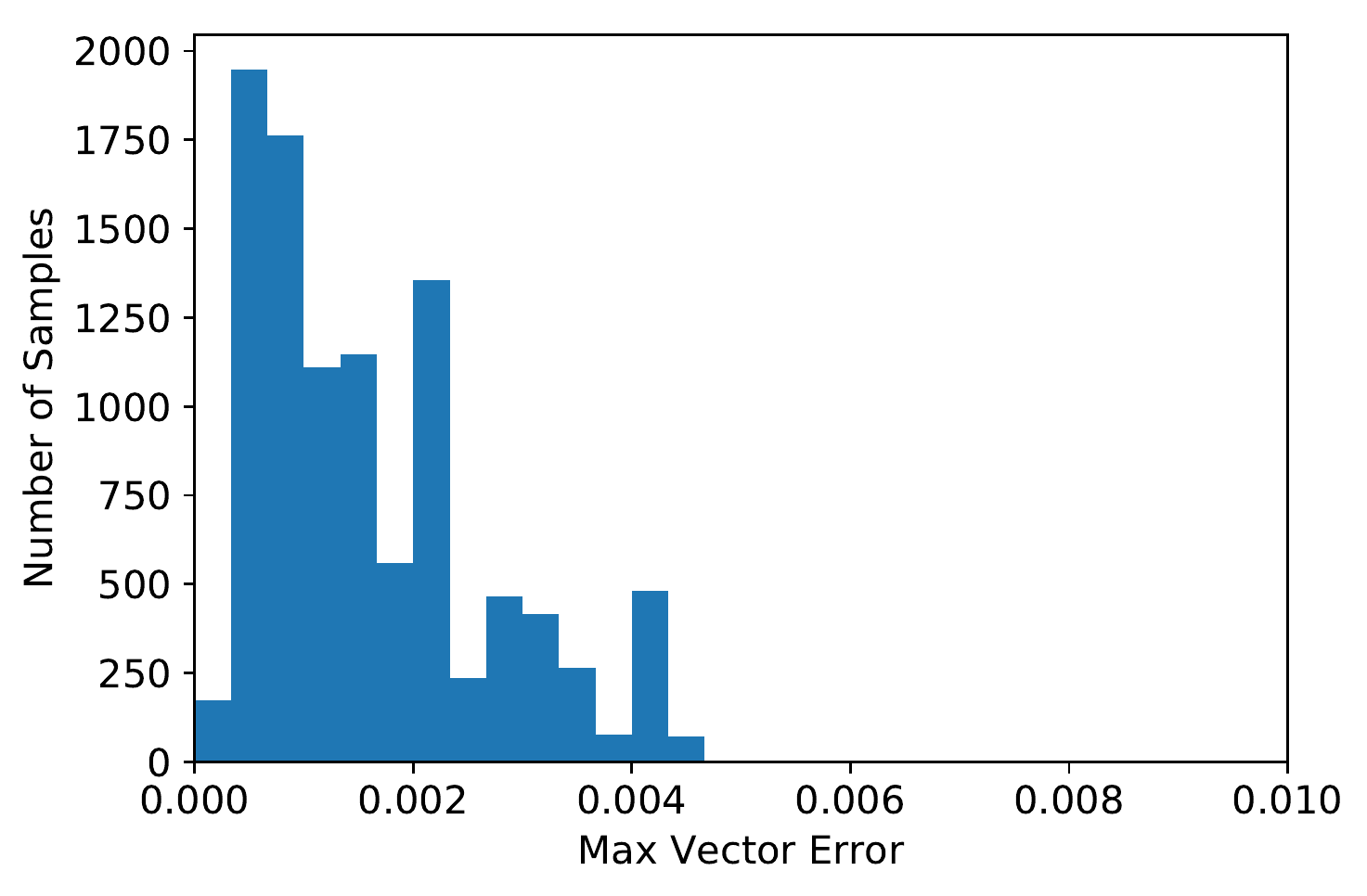}
    \caption{Month-long data set with no clustering: overall prediction error, $\epsilon_{\text{inf}\vec{v}}$, on the training set data. }
    \label{fig:noclust_trainerror}
\end{figure}

\begin{figure}[thb]
    \centering
    \includegraphics[width=\columnwidth]{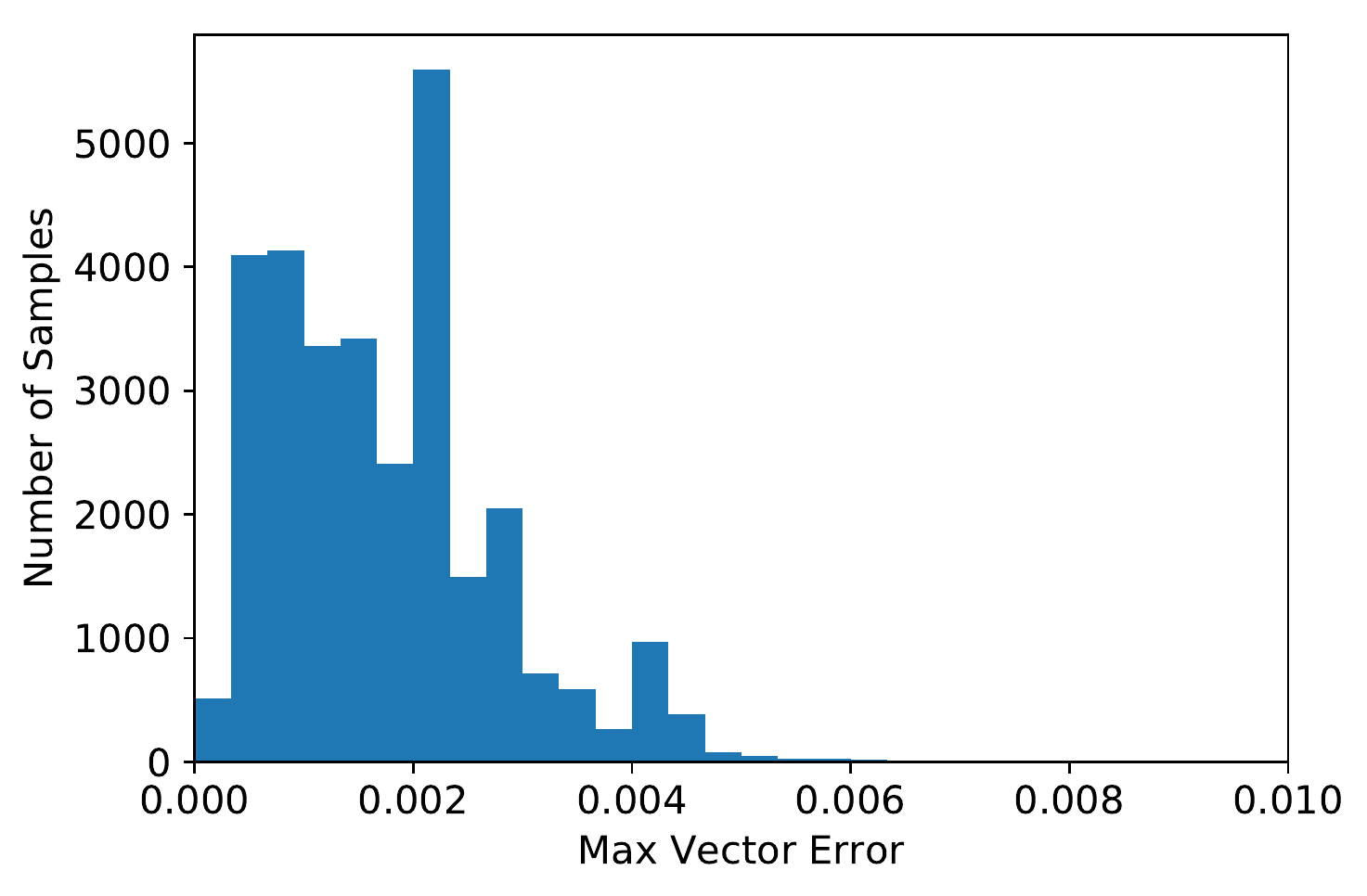}
    \caption{Month-long data set with no clustering: overall prediction error, $\epsilon_{\text{inf}\vec{v}}$, on the testing set data. }
    \label{fig:noclust_testerror}
\end{figure}

The initial method learns only one model for the network and uses no clustering, setting a baseline level of performance. The results are shown in Figures \ref{fig:noclust_trainerror} and \ref{fig:noclust_testerror}. The error distributions are similar between the training and testing sets, though we observe a very small number of extreme values in the test set which stretch the tail of the distribution. Centered around 0.2\% overall prediction error these results do not match the accuracy level of the solver, but for many applications and studies using tools like GridLAB-D this is a sufficient estimate. 

\begin{figure}[thb]
    \centering
    \includegraphics[width=\columnwidth]{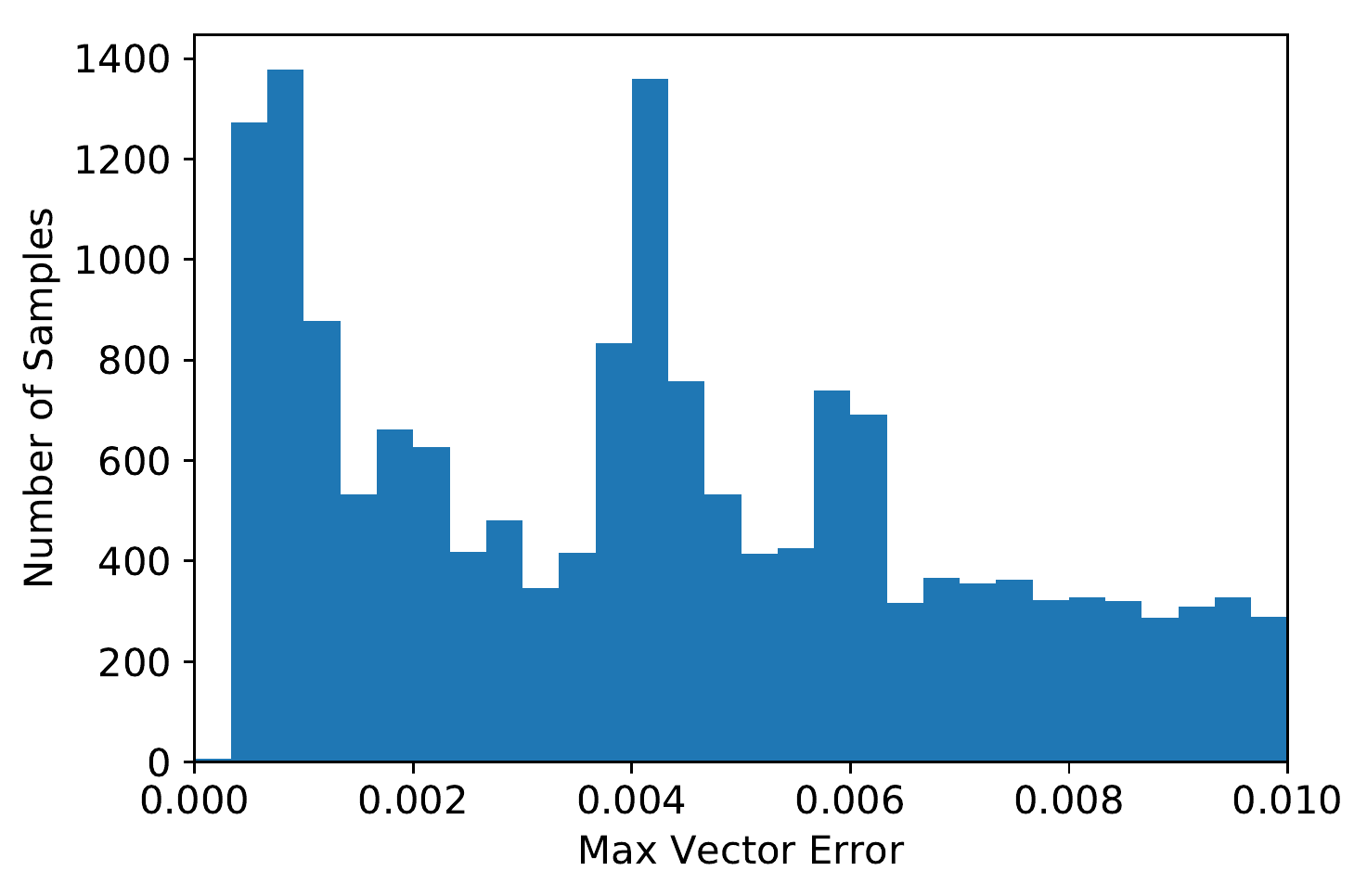}
    \caption{Trimmed histogram of the errors on the test set using 7 cluster models, one for each day of the week. 45.9\% of the samples were excluded for being above this cut-off, and the maximum value was 1.4.}
    \label{fig:7dow}
\end{figure}

\begin{figure}[thb]
    \centering
    \includegraphics[width=\columnwidth]{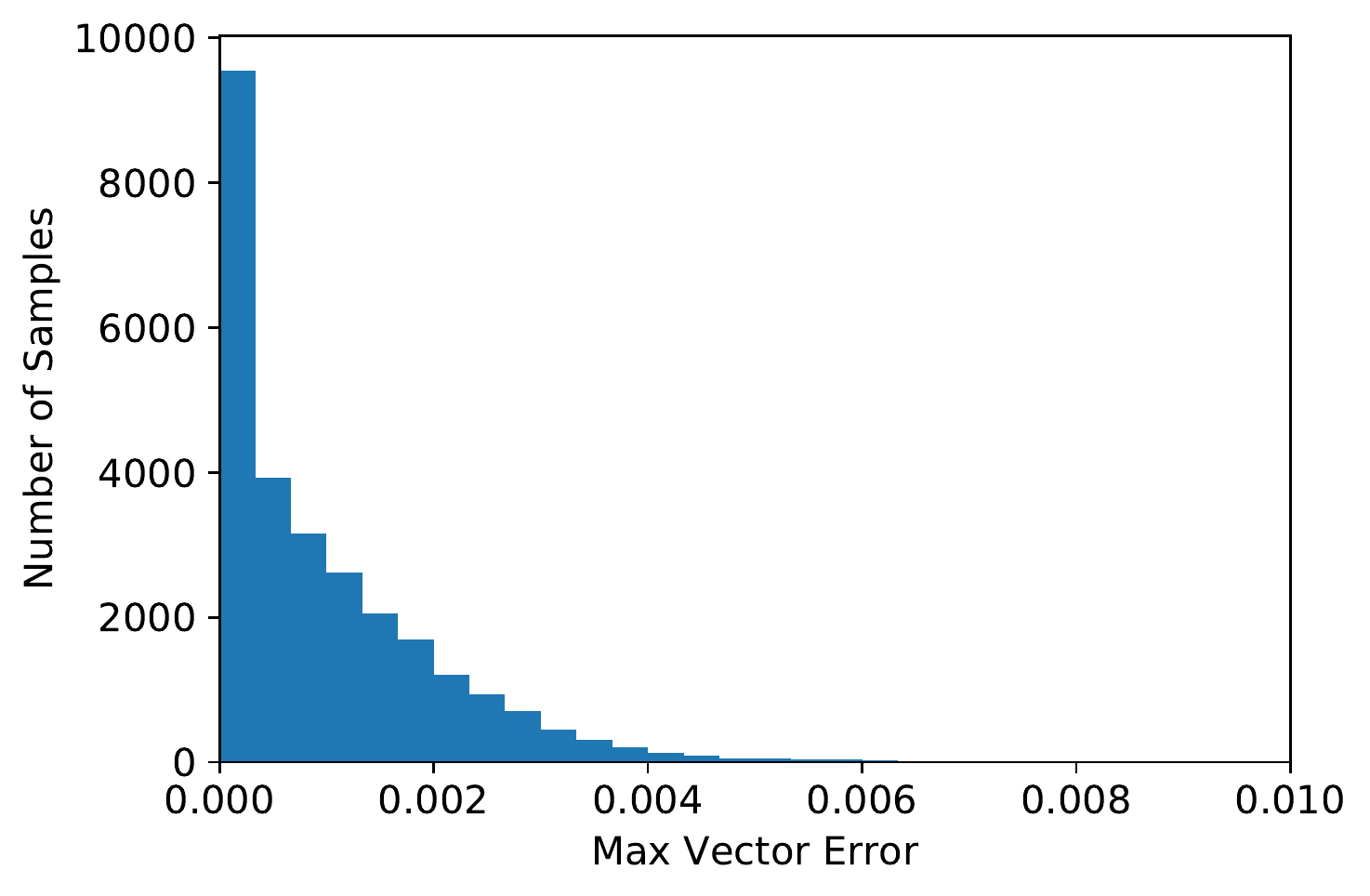}
    \caption{Trimmed histogram of the errors on the test set using 7 K-Means cluster models. 9.7\% of the samples were excluded for being above this cut-off, and the maximum value was 4.3e7.}
    \label{fig:7kmeans}
\end{figure}

\begin{figure}[thb]
    \centering
    \includegraphics[width=\columnwidth]{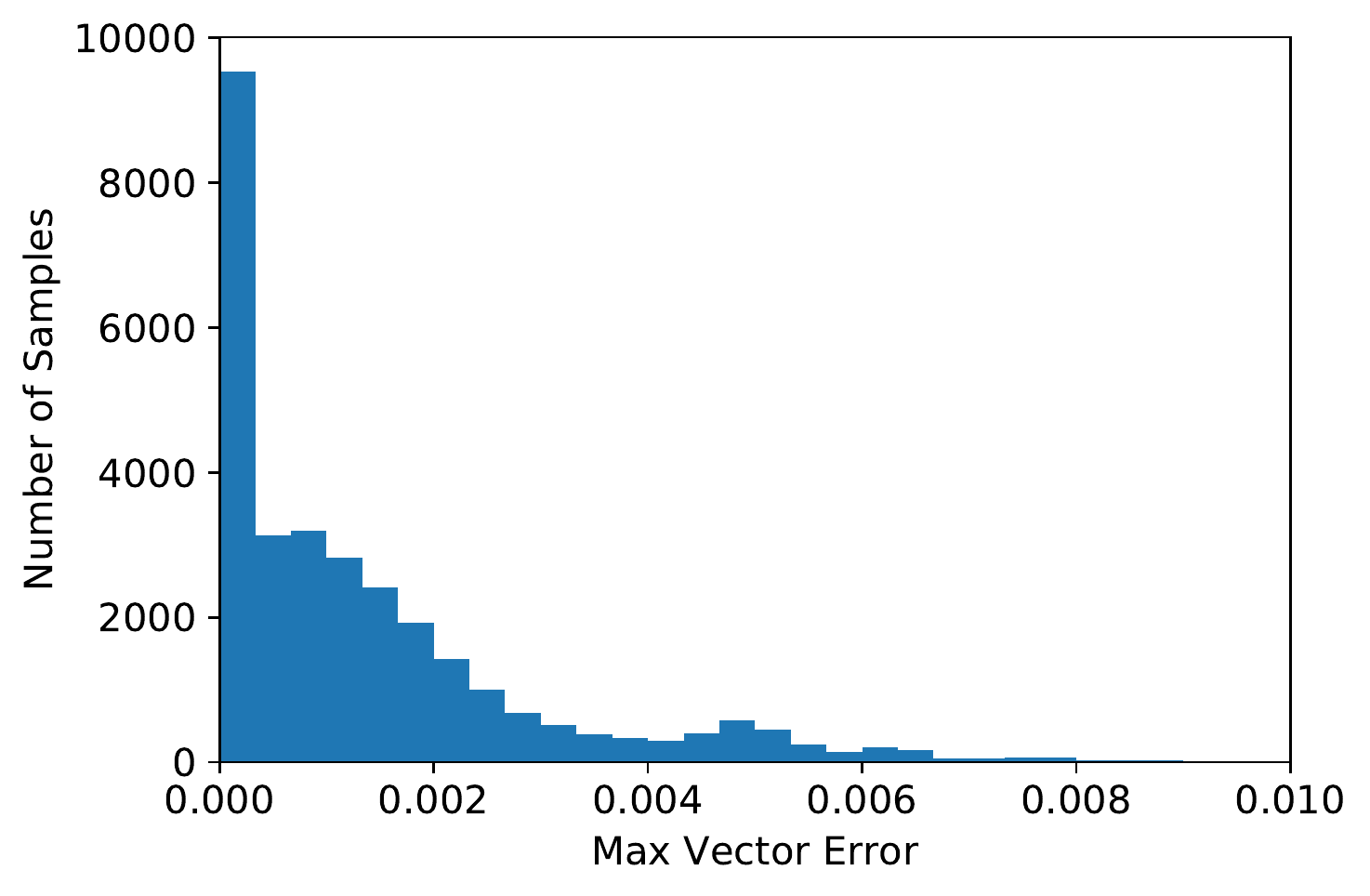}
    \caption{Trimmed histogram of the errors on the test set using 7 Gaussian Mixture cluster models. 0.3\% of the samples were excluded for being above this cut-off, and the maximum value was 125.4. }
    \label{fig:7gauss}
\end{figure}

The time-series of the error shows clear patterns of time dependence in the models performance throughout the day. We capture this by clustering in time using the three previously discussed methods, and the prediction errors for the testing set are presented here: Figure \ref{fig:7dow} shows the error distribution when clustered based on the day of the week, Figure \ref{fig:7kmeans} shows that result for K-Means clustering, and Figure \ref{fig:7gauss} shows that result for Gaussian Mixture Model clustering. Each of these distributions includes a small number of extreme values, and the figures shown here are clipped. For the testing set the three models respectively have 45.9,  9.7, and 0.3\% of values above the cut-off, and maximum errors of 1.4, 4.3e7, and 125.4.

From these figures it is clear that clustering by day of the week gives the least favourable results, significantly worse than the initial cluster-free model from Figure \ref{fig:noclust_testerror}. Both K-Means and Gaussian Mixture are improvements over the original and greatly increase the number of samples in the smallest error bin. Both have error tails, but the K-Means method has a smoother distribution of good predictions. Using 7 clusters is enough to represent the system modes well, and increasing the number of clusters seems to flatten the distribution. Therefore we move forward with the K-Means 7 cluster models for the following experiments. 


\begin{figure}[thb]
    \centering
    \includegraphics[width=\columnwidth]{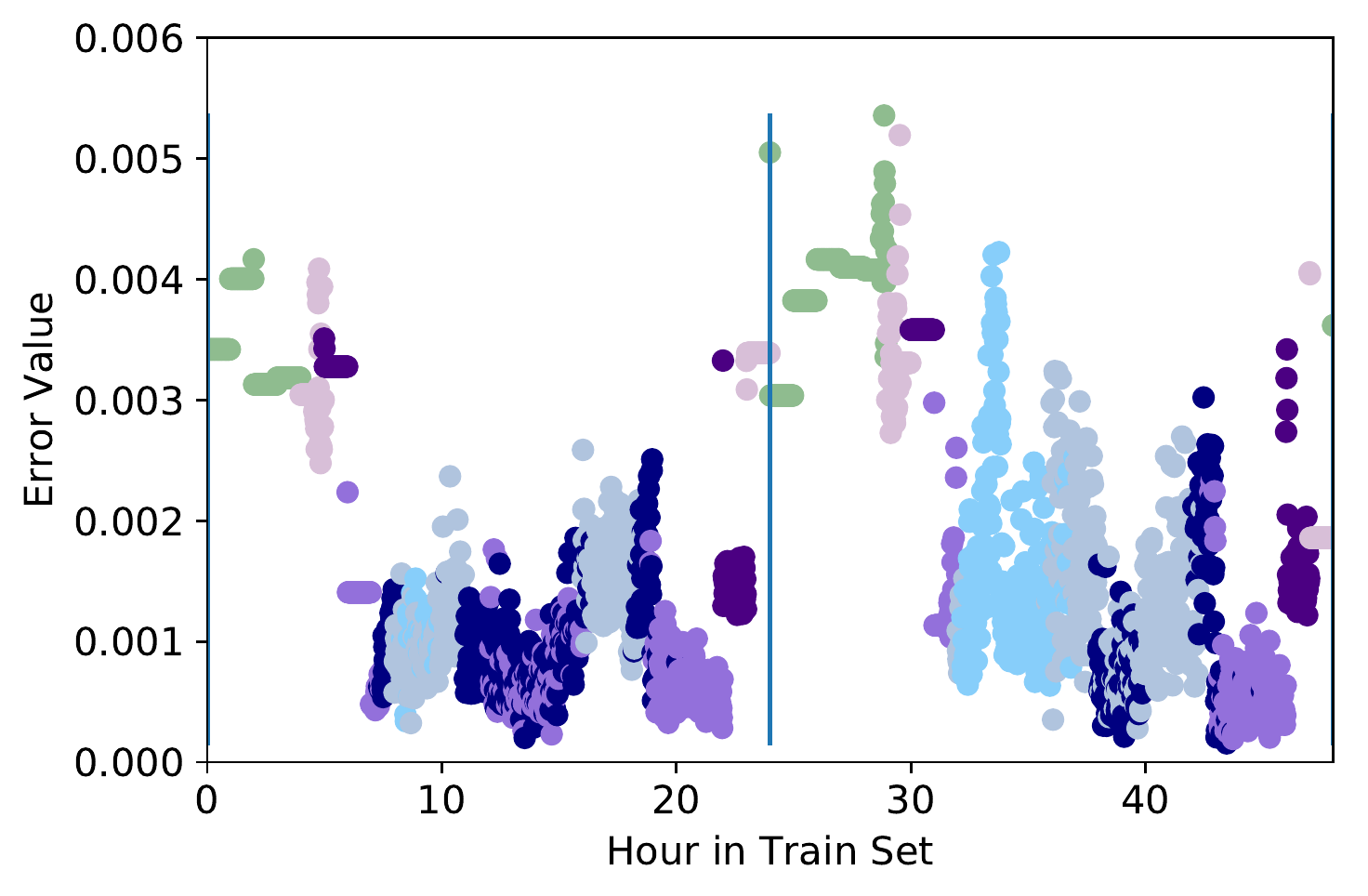}
    \caption{Training set error from the cluster-free model, coloured to show the assignment of points into 7 K-Means clusters. The first two days of the training set data are illustrated here.}
    \label{fig:7clust_kmeans}
\end{figure}

Pictured using the time-series of training set error from Figure \ref{fig:noclust_trainerror}, Figure \ref{fig:7clust_kmeans} shows the distribution of the 7 clusters automatically assigned by K-Means clustering. 

\begin{figure}[thb]
    \centering
    \includegraphics[width=\columnwidth]{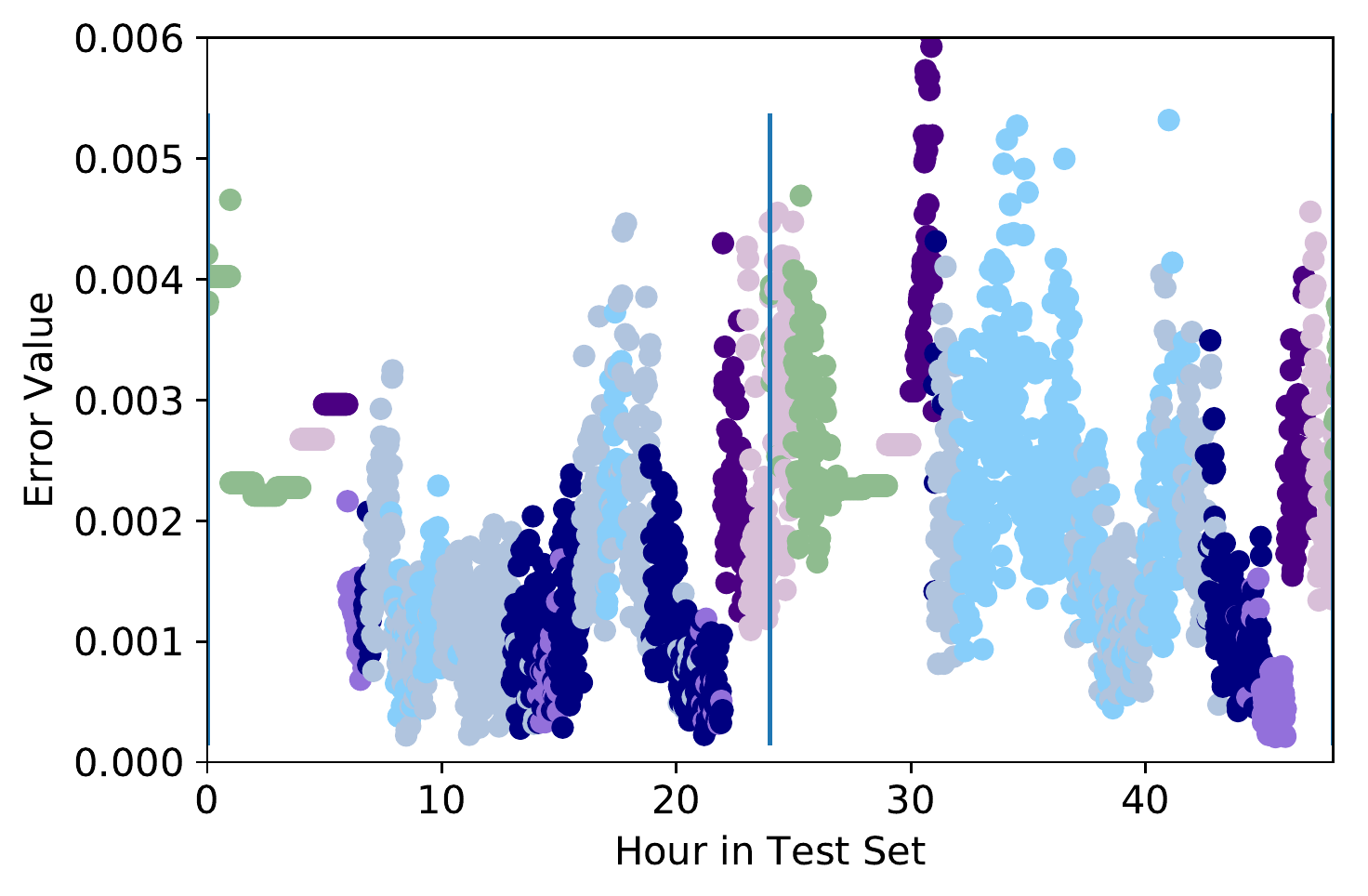}
    \caption{Testing set error from the cluster-free model, as in Figure \ref{fig:7clust_kmeans}, coloured to show the assignment of points into 7 K-Means clusters. Only the first two days of the testing set data are illustrated. These two days of the week align with the two days shown in Figure \ref{fig:7clust_kmeans}, so the differences highlight the complexity of the clustering and system modes.}
    \label{fig:7clust_kmeans_test}
\end{figure}

Investigating the tails of the distribution in Figure \ref{fig:7kmeans} shows that the extreme error results always occur for samples far from their cluster center. If the cluster sample-to-center distance is found for each vector in the training set, we can calculate in what percentile of that distribution the extreme error sample falls. We find this is consistently very high, motivating the Distance Check error parameter described in Section \ref{sec: paramtuning}.


\subsection{Error Parameter Tuning} \label{sec: paramtuning}

\begin{figure*}
\begin{subfigure}[t]{0.45\textwidth}
\centering
    \includegraphics[width=\columnwidth]{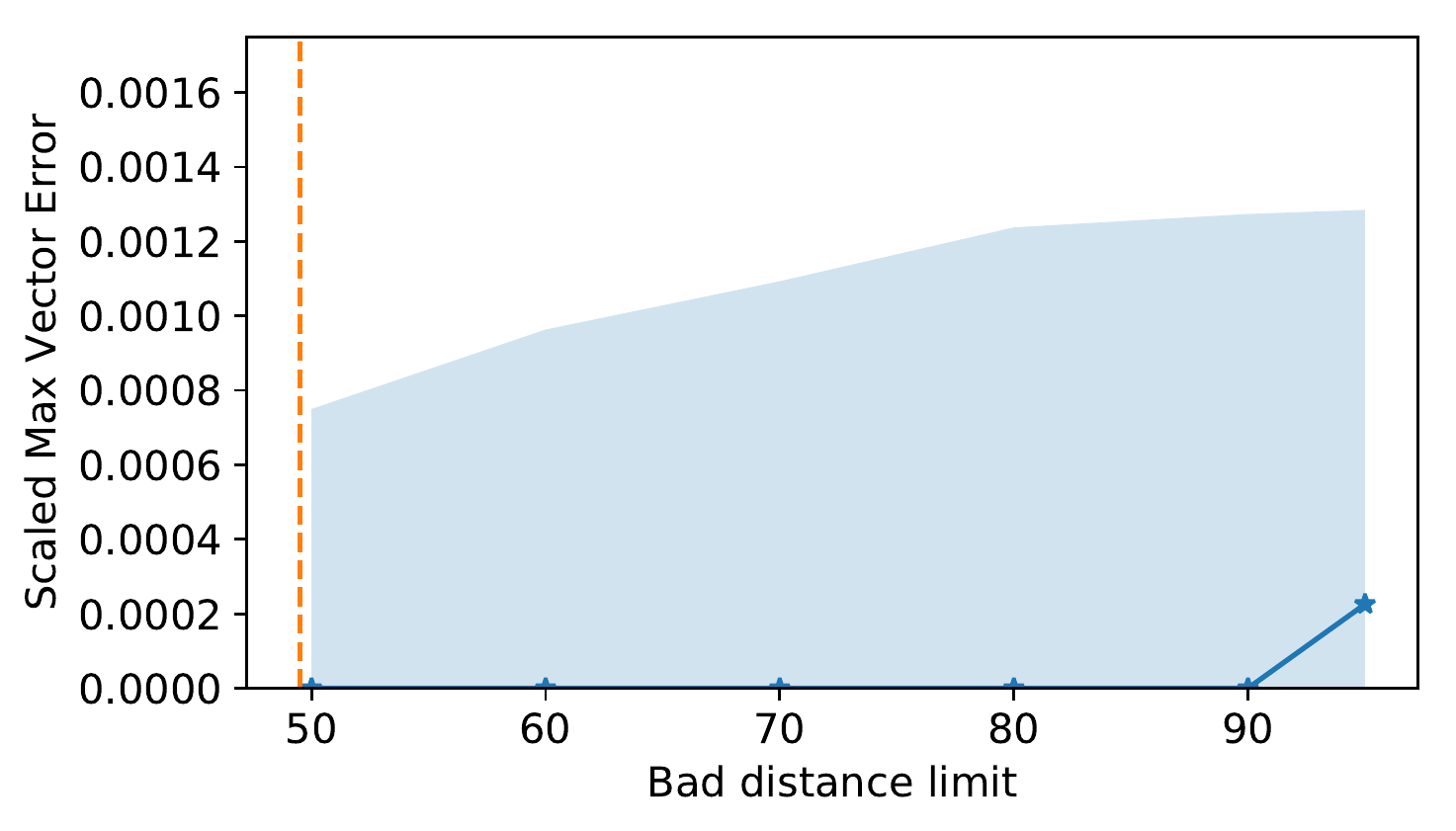}
    \caption{Error based on tuning the Distance Check.}\label{fig:disttuning}
\end{subfigure}
\begin{subfigure}[t]{0.45\textwidth}
\centering
    \includegraphics[width=\columnwidth]{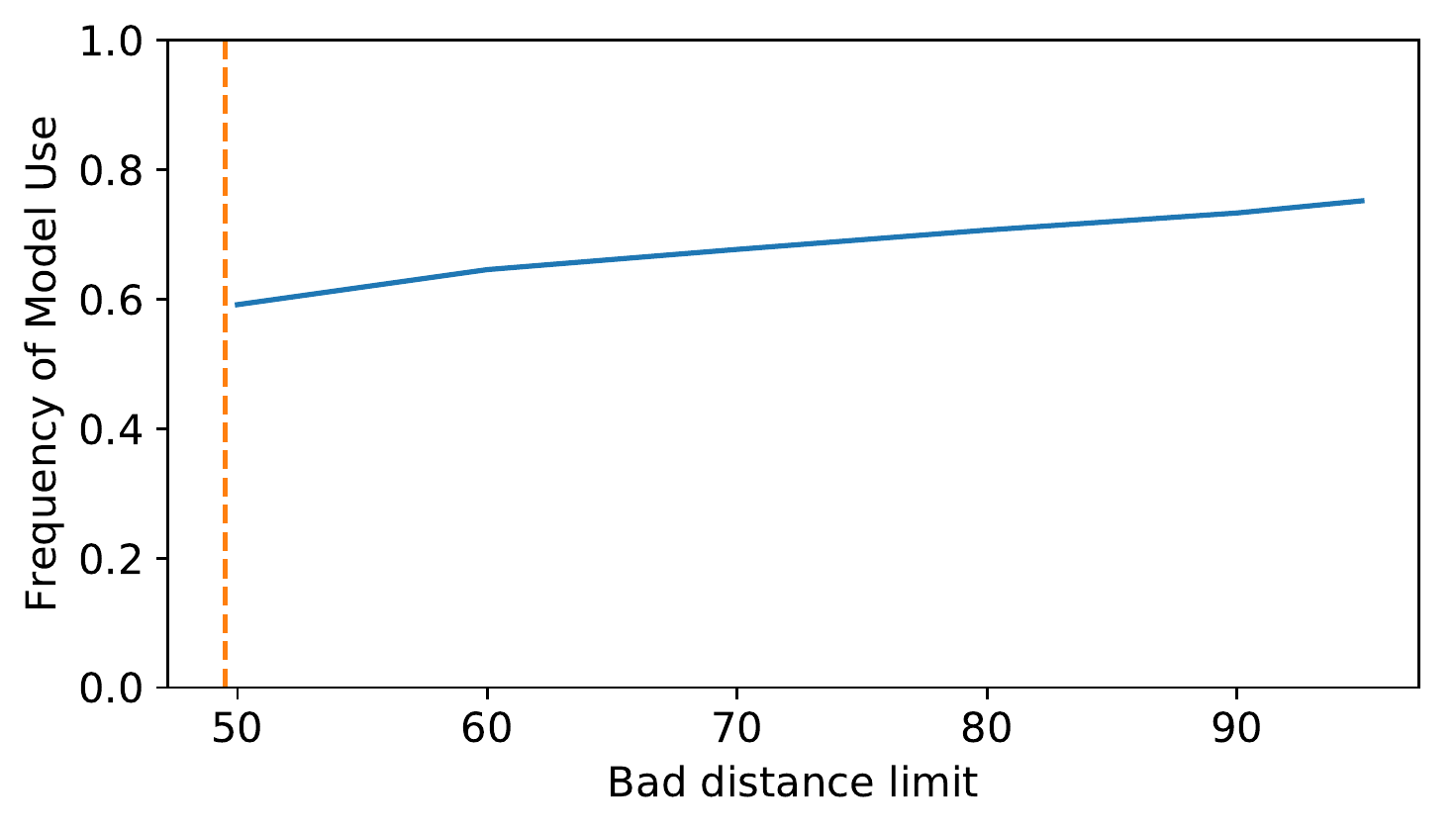}
    \caption{Fraction of times the model was used (rather than the solver) with tuning of the Distance Check.}\label{fig:nummodelcalls_disttuning}
\end{subfigure}
\begin{subfigure}[t]{0.45\textwidth}
\centering
    \includegraphics[width=\columnwidth]{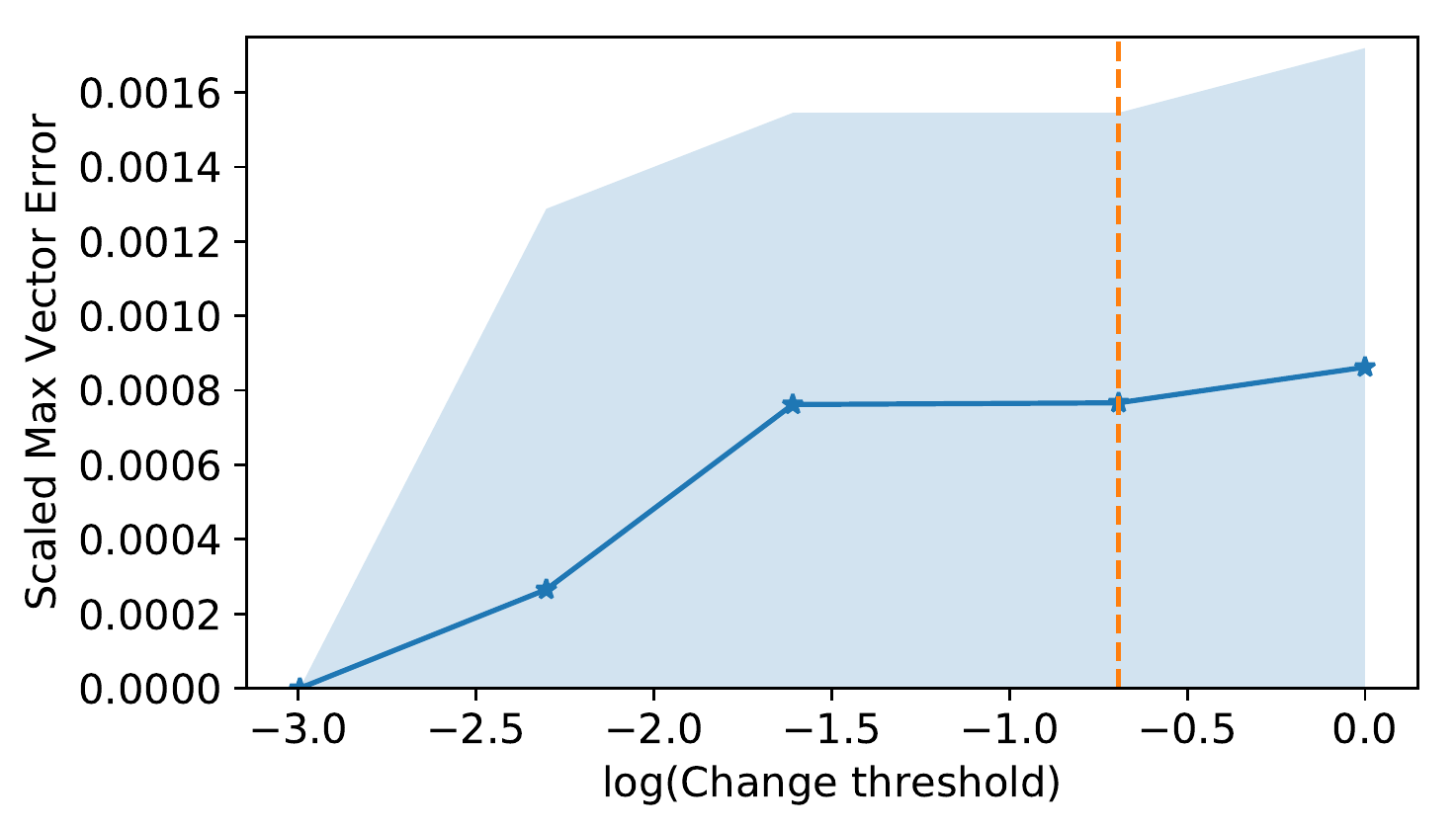}
    \caption{Error based on tuning of the Step Change Check.}\label{fig:changetuning}
\end{subfigure}
\begin{subfigure}[t]{0.45\textwidth}
\centering
    \includegraphics[width=\columnwidth]{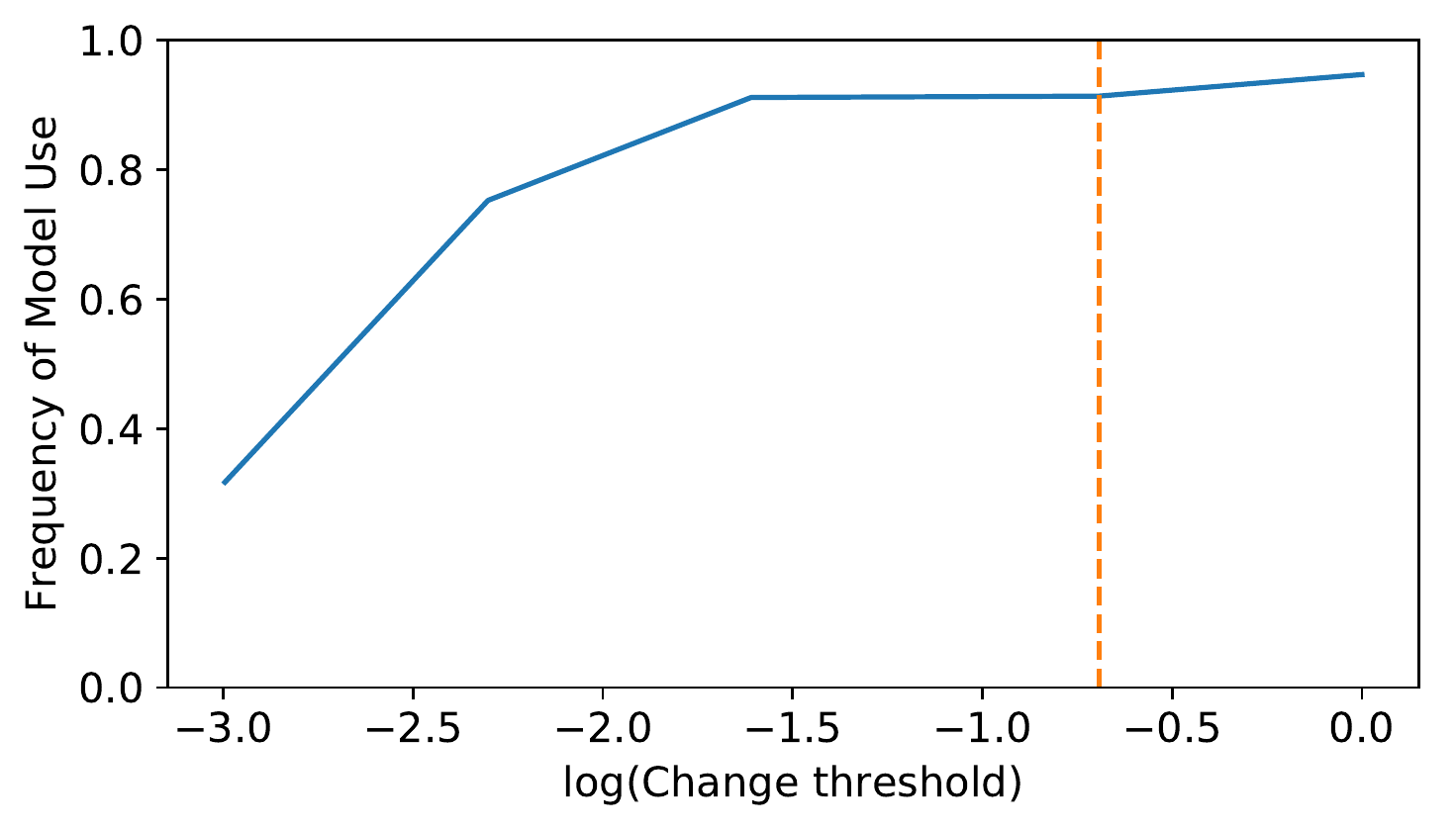}
    \caption{Fraction of times the model was used (rather than the solver) with tuning of the Step Change Check. }\label{fig:nummodelcalls_changetuning}
\end{subfigure}
\begin{subfigure}[t]{0.45\textwidth}
\centering
    \includegraphics[width=\columnwidth]{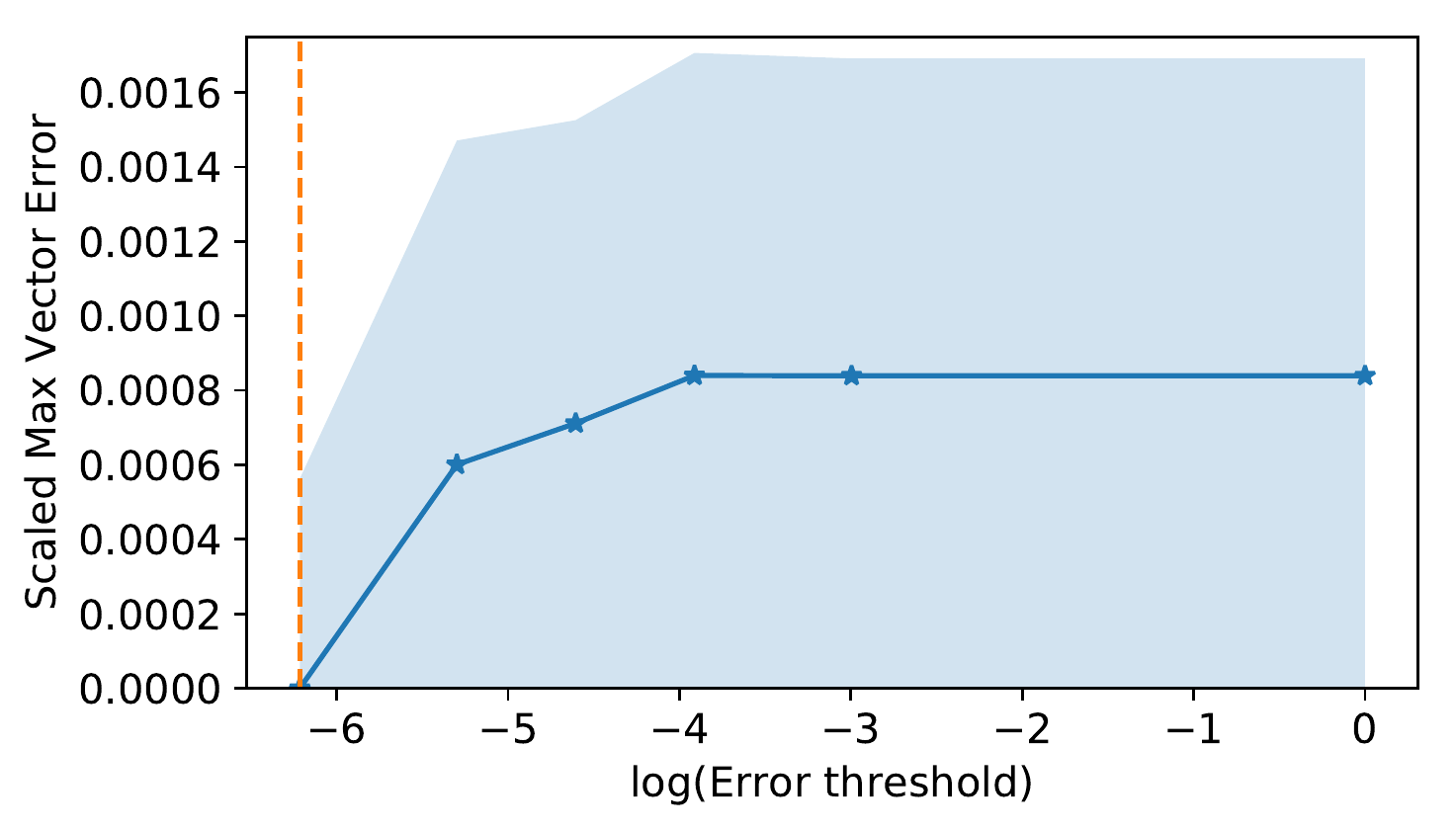}
    \caption{Error based on tuning of the Error Check threshold, with a maximum interval of one hour between checks.}\label{fig:errortuning}
\end{subfigure}
\hfill
\begin{subfigure}[t]{0.45\textwidth}
\centering
    \includegraphics[width=\columnwidth]{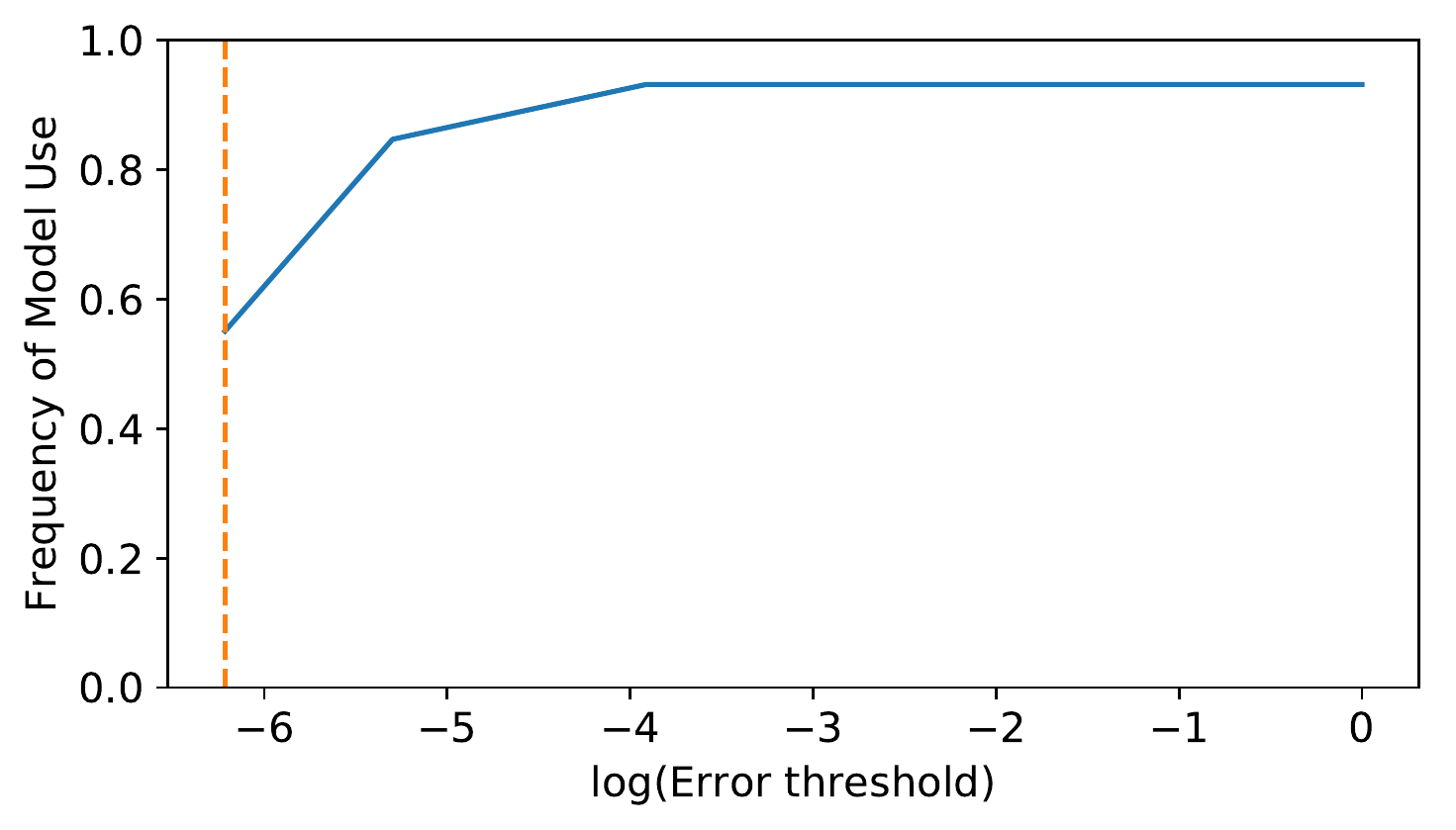}
    \caption{Fraction of times the model was used (rather than the solver) with tuning of the Error Check threshold, with a maximum interval of one hour between checks.}\label{fig:nummodelcalls_errortuning}
\end{subfigure}
\begin{subfigure}[t]{0.45\textwidth}
\centering
    \centering
    \includegraphics[width=\columnwidth]{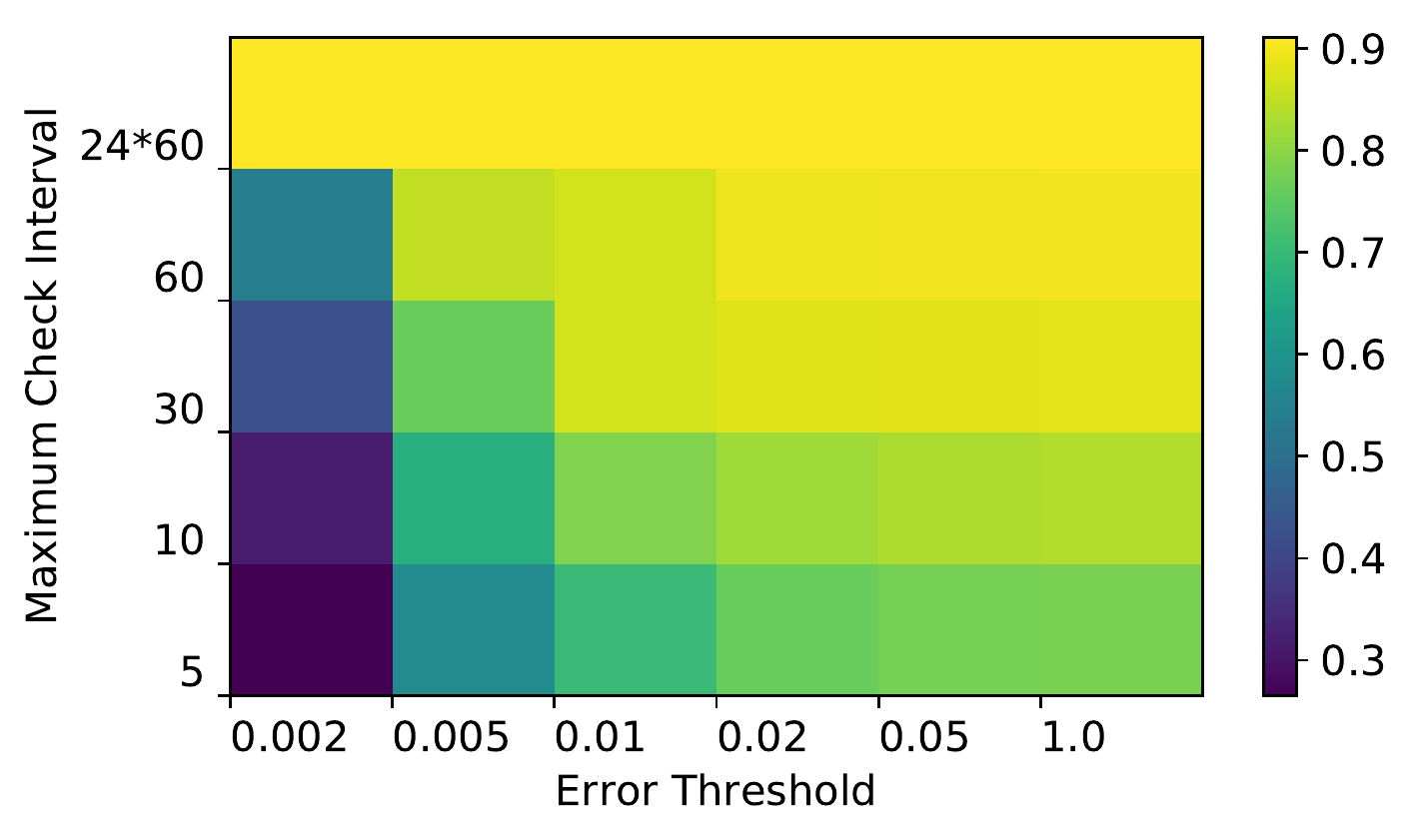}
    \caption{Fraction of test samples for which the model was used in the Error Check sensitivity test.}
    \label{fig:lengths_pcolor}
\end{subfigure}
\hfill
\begin{subfigure}[t]{0.45\textwidth}
\centering
    \centering
    \includegraphics[width=\columnwidth]{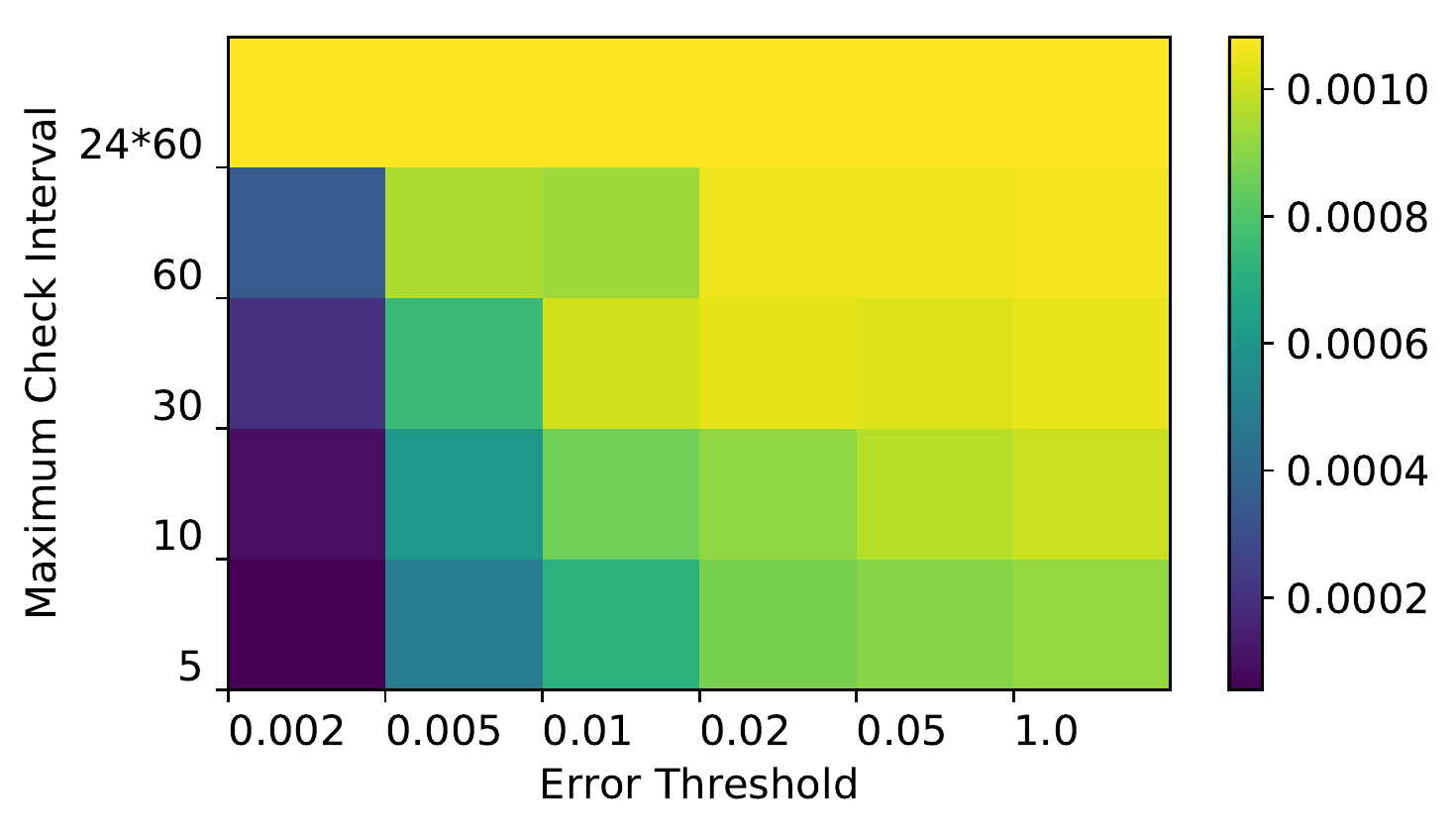}
    \caption{Mean $\epsilon_{\text{inf}\vec{v}}$ error across the Error Check sensitivity test samples.}
    \label{fig:meanerror_pcolor}
\end{subfigure}
\caption{Results of parameter calibration.}
\end{figure*}

Each of the three error parameters was calibrated using the first day of the testing set. The parameters were implemented with a range of values and the prediction errors are shown here in Figures \ref{fig:disttuning}, \ref{fig:changetuning}, and \ref{fig:errortuning}. For each parameter value the error is calculated for every testing sample in the calibration day, and the 25th, 50th, and 75th percentiles of those results are shown by the shaded regions and solid lines. The vertical dashed lines shows the cut-offs after which extreme error values from the tail of the testing set error distribution (orders of magnitude off) start to enter the solution.  
We can also compare the results by how often the solver was called. Figures
\ref{fig:nummodelcalls_disttuning}, \ref{fig:nummodelcalls_changetuning}, and \ref{fig:nummodelcalls_errortuning} show these results.


Comparing between the two sets of figures we see that using the Step Change Check gives the best guarantee against large errors, as we can reach the higher value of approximately 91\% model uses before observing any extreme results.  In that case the solver was avoided more than 90\% of the time while achieving a median error of less than 0.1\%. This reduces the simulation time significantly by minimizing the number of solutions requiring the full solver. Compared with the Error Check, it reaches the same model count and median error at approximately the same time, but is more robust to outliers. It is also noticeable that the Distance Check gives a very low median error, though some of this may be attributable to its lower number of model uses.


Some combination of these checks is optimal. Taking from each of their performance highlights for this network we set the Step Change Check threshold to 20\% (-1.6 on the log scale plots) and use an Error Check to try to take advantage of its steeper rise in model uses. 

The value of the Error Check threshold should depend on how frequently a check is forced. Without requiring occasional solves to update the error metric, the metric value can become out of date; this may have contributed to the extreme errors included in this test, as the Error Check value was not updated frequently enough to catch those events.

To find the optimal parameters, a sensitivity test was conducted over the values of the Error Check. No Distance Check was used, and the Step Change Check threshold was implemented at 20\%. The sensitivity results are shown in Figures \ref{fig:lengths_pcolor} and \ref{fig:meanerror_pcolor}, comparing the parameters in terms of their time savings and accuracy. 
Analyzing the maximum test error rather than the mean showed a very similar distribution of results to Figure \ref{fig:meanerror_pcolor}. Together these results point to the best solution of parameters to achieve high accuracy and avoid using the solver: set the maximum interval between checks at 60 minutes and use 0.01 or 1\% error as the threshold for when to resume using the solver.

\subsection{Final Model} \label{sec:final_results}
\begin{figure}[htb]
    \centering
    \includegraphics[width=\columnwidth]{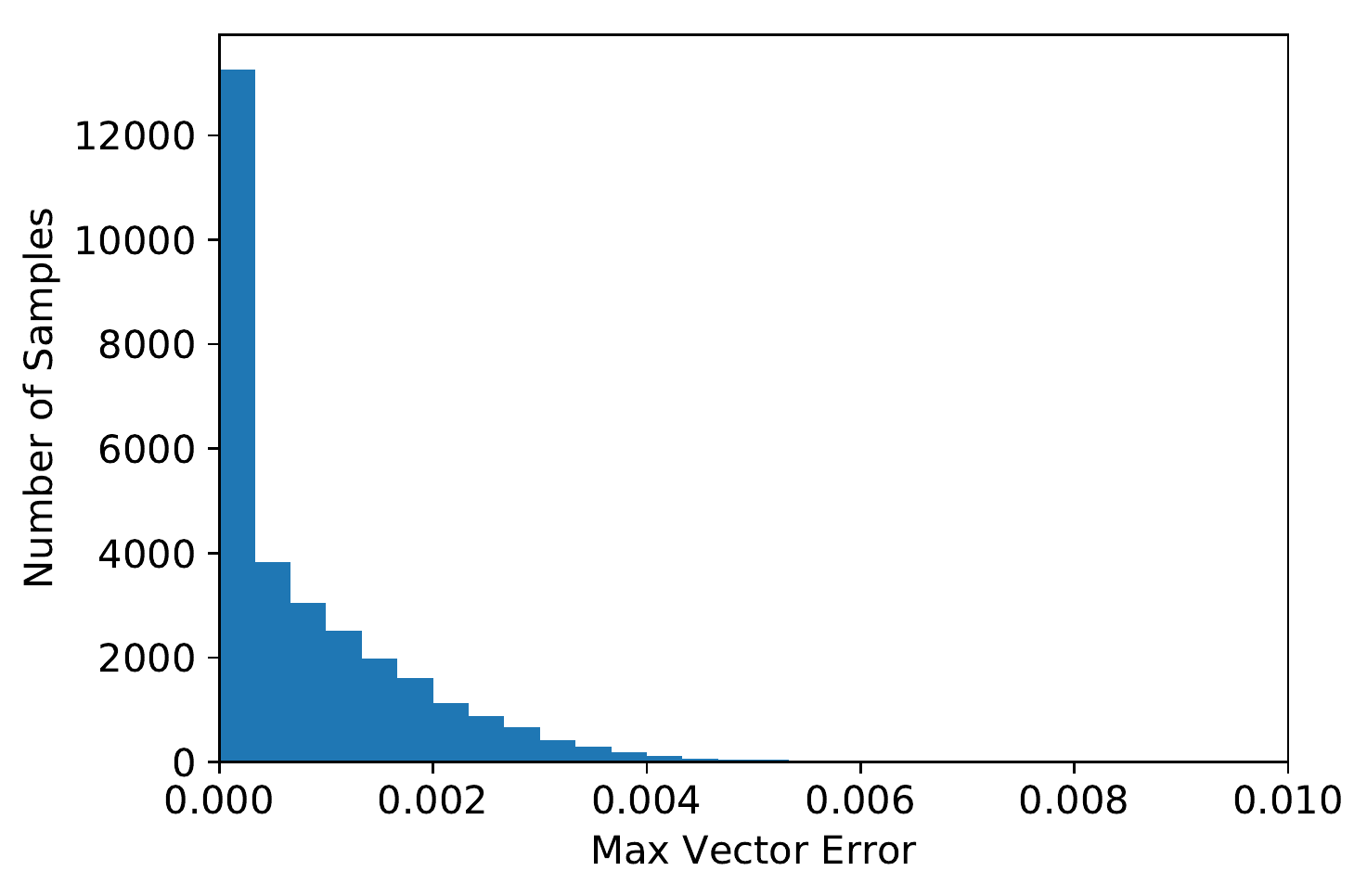}
    \caption{Trimmed histogram of the overall prediction errors on the testing set using the final model. 0.05\% of the samples were clipped and the maximum value was 0.0168.}
    \label{fig:finalerror}
\end{figure}

The final parameters of the model for this network are outlined in Table \ref{tab: final_params}. This model was applied to the full test set. The test error results shown in Figure \ref{fig:finalerror} reveal a smooth distribution which grows steeply and has a high peak of near-zero error estimates.

\begin{table}[thb]
\centering
\caption{Final model design.}
\begin{tabular}{l|l}
\hline \bf Model Choice & \bf Value  \\ \hline
Train set length & 1 week \\
Test set length & 3 weeks  \\
Clustering method & K-Means \\
Number of clusters & 7 \\ 
\hline \bf Error Parameter & \bf Value \\ \hline
Distance Check & None \\
Error Check threshold & 1 \% \\
Max interval between Error Checks & 60 min \\
Step Change Check threshold & 20 \% \\
\hline
\end{tabular}
\label{tab: final_params}
\end{table}

\section{Discussion} \label{sec:disc}

There are two main comparison points for these results: one, the cluster-free initial approach; and two, the model-free solver-only approach. 

\begin{table}[thb]
\centering
\caption{Comparison of model results.}
\begin{tabular}{l|c|c|c}
\hline \bf Metric & \bf Solver & \bf Initial & \bf Final \\ \hline
Avoided solves & 0 & 100\% & 86.7\% \\
Median test $\epsilon_{\text{inf} \vec{v}}$ & 0 & 0.0014 & 0.00049 \\
Max test $\epsilon_{\text{inf} \vec{v}}$& 0 & 0.0094 & 0.0168\\
Clipped above 0.01 & 0 & 0 & 0.05\% \\
\hline
\end{tabular}
\label{tab: results}
\end{table}

\begin{figure}[htb]
    \centering
    \includegraphics[width=\columnwidth]{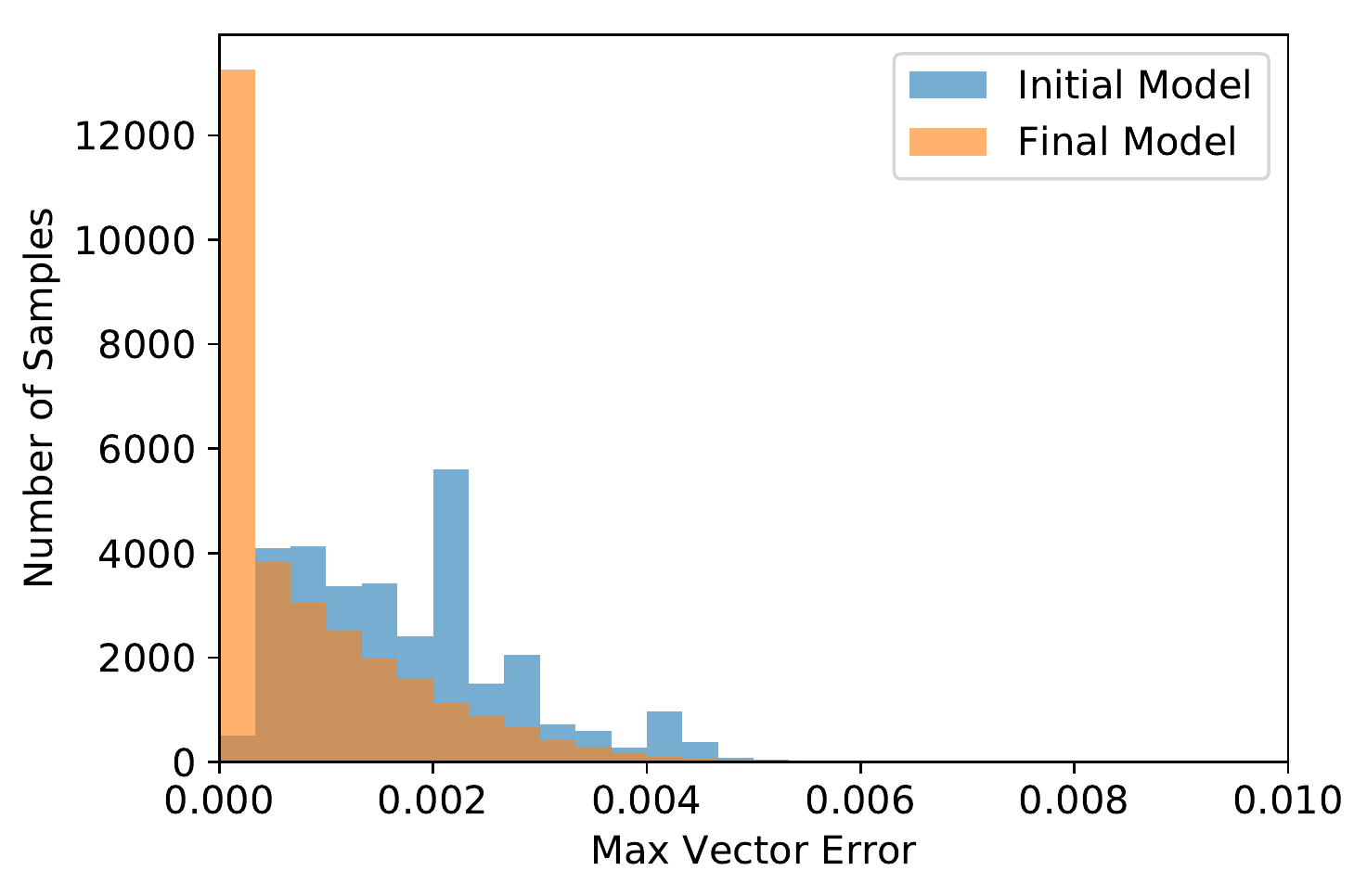}
    \caption{Histograms of the overall prediction error for the initial and final models on the full testing set, highlighting the improvement in the error distribution. This is a combination of Figures \ref{fig:noclust_testerror} and \ref{fig:finalerror}.}
    \label{fig:doublehist}
\end{figure}

The three simulations are compared in Table \ref{tab: results}. Comparing the test error distributions of the initial and final models in Figure \ref{fig:doublehist}, it is easy to see the impacts of the model adjustments.  

Compared with the first clustering results in Figure \ref{fig:7kmeans}, the distribution tail of high errors has been virtually eliminated: only 0.05\% of the test samples see errors higher than 1\% and none of those are higher than 1.68\%. 

The final model calls on the solver for 13.3\% of the samples so it is slower than the initial model, but that is balanced by the improvement in the median test error. With a median overall prediction error, $\epsilon_{\text{inf} \vec{v}}$, of just under 0.05\%, the final model is much more accurate. 

This error compared to the solver is small enough for the method to be used in many power system modeling applications, especially when it allows the simulation to avoid expensive full solves 86.7\% of the time. 

Making more conservative choices for each of the error parameters would decrease this error result even further and target the tail of the distribution, but that would also decrease the utility of this approach by requiring more frequent use of computationally expensive solutions. The choice for this trade-off is certainly problem dependent. 

A limitation of this method is that it gives no hard guarantees on the model error. Without using the solver it is impossible to exactly estimate the error for each sample. Even with conservative safe-guards or applying the model only to samples in areas where confidence in the model is very high, this may pose a problem for applications that are highly sensitive to solver error. 

Tuning these error parameters will likely be different for each network, and the same results cannot be guaranteed if the values selected here in Table \ref{tab: final_params} are naively applied to new systems. This is a limitation as modeling software tools that integrate this approach will either need to choose conservative defaults or include a mechanism early in the simulation to perform training and calibration. 


\section{Future Work} \label{sec:future}

We plan to pursue many different avenues of future work expanding on this idea.

We will deploy this methodology in GridLAB-D using an existing program to boost that tools performance for use-cases that are important to California's renewable energy policies. In doing so we will apply the method to larger systems including the IEEE 8500 node test model, the North American Taxonomy Feeder models \cite{Schneider2009ModernReport}, and real system models provided by utilities to evaluate the method's performance impact on analyses they perform on their distribution systems.

Future work will include automating the error tuning mechanism and finding more data-driven ways to predict each samples test error, for example by learning to predict the NR solver steps required for each input and using that metric as a decision making tool.

We also plan to study how this method could be adapted using periodic re-clustering and re-training of the models for longer time-series simulations where the system evolves slowly over time, such as growing solar photovoltaic penetration or battery storage. The training set can change dynamically as new operating modes emerge, and the process for choosing which samples to keep will become part of the research question.

Another extension will be to consider more clustering methods and study how the clustering results represent phenomena in the data set or the solver.

\section{Conclusion} \label{sec:conc}

A data-driven power flow solution method was developed for a typical distribution system model. Its implementation was shown to reduce the computation time required for a large-scale quasi-steady state power flow simulation. By coupling the solver with a data-driven approach, the usage of the physics-based solver was reduced by 86.7\% and the median overall prediction error achieved was 0.049\%. 

Further investigation of the methodology will include a variety of larger distribution networks as well as longer time-series demonstrations and periodic re-training of the model. To improve the speed and efficiency of utility planning studies and the evaluation of operation strategies developed by researchers and electrical utility companies, this method will be implemented in GridLAB-D. 


\section*{Acknowledgements} The authors want to thank Lily Buechler for her helpful advice and feedback throughout this work. Thanks also to the Grid Integration Systems and Mobility (GISMo) team at SLAC National Accelerator Laboratory for their support, feedback, and encouragement. This work was funded by the California Energy Commission under grant EPC-17-046. SLAC National Accelerator Laboratory is operated for the US Department of Energy by Stanford University under Contract DE-AC02-76SF00515.

\bibliographystyle{ieeetr}
\bibliography{references}

\end{document}